\UseRawInputEncoding
\documentclass[aps,twocolumn,nofootinbib]{revtex4-1}
\usepackage{amsmath}
\usepackage{amsfonts}
\usepackage{amssymb}
\usepackage{graphicx}
\usepackage{bm}
\usepackage{color}
\usepackage{ulem}
\def\calO {{\cal O}} 
\usepackage{amsmath}
\usepackage{amsfonts}
\usepackage{amssymb}
\usepackage{graphicx}
\usepackage{bm}
\usepackage{color}
\usepackage{ulem}
\def\calO {{\cal O}} 
 \usepackage{epsfig}
\usepackage{graphicx}
\usepackage{chngcntr}
\counterwithout{figure}{section}
\usepackage{rotating}
\usepackage{amssymb}
\usepackage{amsmath}
\usepackage{physics}
\usepackage{multirow}
\usepackage{float}
\usepackage{caption, subcaption}

\begin{document} 

\title{\bf Conformal Bootstrap Signatures of the Tricritical Ising Universality Class}

\author{Chethan N Gowdigere}
\author{Jagannath Santara}
\author{Sumedha}
\affiliation{School of Physical Sciences, National Institute of Science Education and Research, Jatni - 752050,India}
\affiliation{Homi Bhabha National Institute, Training School Complex, Anushakti Nagar, Mumbai 400094, India}

\date{\today}

\begin{abstract}

We study the tricritical Ising universality class using conformal bootstrap techniques.  By studying bootstrap constraints originating from multiple correlators on the CFT data of multiple OPEs, we are able to determine the scaling dimension of the spin field $\Delta_\sigma$ in various non-integer  dimensions $2 \le d \le 3$.  $\Delta_{\sigma}$ is connected to the critical exponent $\eta$ that governs the (tri-)critical behaviour of the two point function via the relation, $\eta = 2 - d + 2 \Delta_{\sigma}$. Our results for $\Delta_\sigma$ match with the exactly known values in two and three dimensions and are a conjecture for  non-integer dimensions.  We also compare our CFT results for $\Delta_\sigma$ with  $\epsilon$-expansion results, available up to $\epsilon^3$ order.  Our techniques can be naturally extended to study higher-order multi-critical points.

\end{abstract}

\maketitle

\section{Introduction}

The consequences of the conformal hypothesis which posits conformal invariance to the behaviour of physical systems at criticality in addition to scale invariance are most far-reaching in two dimensions, where the conformal symmetry is the infinite dimensional Virasoro algebra.  The seminal work of Belavin, Polyakov and Zamoldchikov \cite{Belavin:1984vu, DiFrancesco:1997nk} resulted in the discovery of a whole class of two dimensional conformal field theories (CFTs) viz. the Virasaro minimal models. Each of these models describes a universality class and the exact knowledge of the scaling dimensions of the operators amounts to a derivation of the critical exponents purely from conformal invariance. The Ising model, the tricritical Ising model, the three and four state Potts models in two dimensions were thus exactly solved thirty years ago. The infinitude of conformal symmetry made two dimensions rather special and the use of conformal field theories to study critical phenomena was restricted to two dimensions alone.

Recent times have seen a breakthrough in doing the same for three and other dimensions. Even though the conformal group is finite dimensional now, technical advances made in the explicit computation of (global) conformal blocks\, (first in \cite{Dolan:2000ut, Dolan:2003hv, Dolan:2011dv} and later in \cite{Hogervorst:2013sma, Kos:2013tga, Kos:2014bka}), resulted in astounding progress (see \cite{Poland:2018epd} for a recent review) and has provided the most precise values \cite{Rattazzi:2008pe, Rychkov:2011et, ElShowk:2012ht, El-Showk:2014dwa} for the critical exponents of the three dimensional Ising model.  Rychkov et al analyzed the restrictions imposed by conformal invariance and found that the conformal field theory corresponding to the Ising universality class sits on the boundary of the allowed region at a kink-like point in the space of scaling dimensions of the only two relevant operators. Furthermore, remarkably, they could extend this analysis to all non-integer dimensions between two and four and showed that even here, the theory corresponding to the Ising universality class is always located at a kink-like point \cite{El-Showk:2013nia}.  Hitherto \cite{Kos:2015mba, Kos:2016ysd, Bobev:2015vsa, Gliozzi:2013ysa, Nakayama:2016cim, Gopakumar:2016wkt}, only critical points have been studied using conformal field theory methods.  Other critical points (tricritical, multi-critical) also have conformal symmetry and here we use conformal field theory techniques to study tricritical points. In this paper, we look at the tricritical Ising point in two and higher dimensions and show that just like the Ising critical point, the tricritical Ising point can also be recognised by its special signatures in the space of scaling dimensions of the appropriate operators.

A tricritical point is a fixed point where three critical lines and a line of first order transitions meet (or, a point where three coexisiting phases simultaneously become critical)\cite{Griffiths:1970zza}.  Quantum scalar field theory with a $\phi^6$ interaction provides one realisation of this universality class. Here, there is a line of Ising critical points which end in a higher order critical point namely the tricritical point which has a different set of critical exponents.  These two critical behaviours have different upper critical dimensions; three for the tricritical Ising point and four for the Ising critical point\,\cite{lawrie}. Due to this,  field theory and renormalization group based techniques such as the $\epsilon$-expansion have had limited success\,\cite{Lewis:1978zz, Hager:2002uq}. CFT techniques avoid the flow and only study the fixed point which is where  the extra scaling and conformal symmetries are present. Yet, so far, 
a CFT study of the tricritical point has not met with as much success 
as the critical point, in dimensions other than two. 

The paper is organised as follows.  We first start with a brief review of CFT,  in section two.  We recall  prime principles of CFT, the essential details of the conformal bootstrap program and some key results obtained so far in the bootstrap program that we need for our analysis of the tricritical CFT.  In section three, we present the bootstrap analysis that allows us to compute the tricritical exponent $\eta$ for all non-integer dimensions between two and three. In the same section, we compare with the best available $\epsilon$-expansion results for $\eta$. In section four, we present further bootstrap analysis pertaining to multiple universality classes and multicriticality.  We conclude the paper in section five with a summary of results and future directions.

\section{Brief Review of CFT}
We first provide a quick overview of the prime principles of CFT:  CFT data, the operator product expansion (OPE), unitarity constraints, crossing symmetry constraints.  A CFT is specified (partially) by a list of local primary operators a.k.a scaling operators (each primary operator  is a primary under the global conformal symmetry and is specified by its scaling dimension and spin). The scaling dimension of a local operator in an unitary CFT is bounded from below depending on it's spin.   A product of two such local operators is expandable in terms of all the local operators\,\cite{Wilson:1969zs, Kadanoff:1969zz}; this is known as the operator product expansion
\begin{equation}
\phi_1(x_1)\phi_2(x_2)=\sum_{\calO} \lambda_{12 \calO}  C(x_{12}, \partial_{x_2}) \calO(x_2)\,,
\end{equation}
where $C(x_{12}, \partial_{x_2})$ is fully determined by conformal invariance and $\lambda_{12 \calO}$ is referred to as an OPE co-efficient, which is real in an unitary CFT.  The set of local primary operators and the set of OPE co-efficients  are together referred to as CFT data, which completely specifies a local CFT. 

Critical exponents are encoded in the scaling dimensions of only a few low-lying primary operators (the relevant ones). There are only two relevant operators in the Ising CFT,  while there are four relevant operators in the CFT of the tricritical Ising point\,\cite{lawrie1}.  Of the four, one needs only three to define the critical exponents of the tricritical Ising universality class. In two dimensions these exponents are known exactly\,\cite{Belavin:1984vu, DiFrancesco:1997nk};  the CFT is the  second Virasaro minimal model consisting of six primary operators (while the Ising CFT is the first Virasaro minimal model consisting of three primary operators).

In a CFT, a four point function, shown here for  scalar operators, is mostly fixed by conformal invariance,  except for an arbitrary function of the two independent cross-ratios $u=\frac{x^2_{12}\,x^2_{34}}{x^2_{13}\,x^2_{24}}$ and $v=\frac{x^2_{14}\,x^2_{23}}{x^2_{13}\,x^2_{24}}$ with $x_{12} \equiv | \vec{x}_1 - \vec{x}_2 | ..$:
\begin{equation}
\langle\phi(x_{1})\phi(x_{2})\phi(x_{3})\phi(x_{4})\rangle
\sim G(u,v).
\end{equation}
Using OPE's one can evaluate the four point function in terms of the CFT data 
\begin{equation}
G(u,v)= \sum_\calO \lambda_{12 \calO}\lambda_{34 \calO} {G_\calO(u,v)}\,.
\end{equation}
where the summation is over the primary operators that occur in both the $\phi_1 \times \phi_2$ and the $\phi_3 \times \phi_4$ OPE's and ${G_\calO(u,v)}$ is the (global) conformal block. This evaluation of the four-point function can be done in two different ways and requiring that both ways agree 
\begin{equation} \label{bse}
 \sum_\calO \lambda_{12 \calO}\lambda_{34 \calO} {G_\calO(u,v)}\,= \sum_{\calO'} \lambda_{14 \calO'}\lambda_{23 \calO'} {G_{\calO'}(u,v)}\,
 \end{equation}
puts a constraint on the CFT data which is referred to as the bootstrap equation or also as the crossing symmetry constraint. The conformal bootstrap program employs these equations \eqref{bse} (every four point function provides one equation) to search for CFTs by successively constraining the CFT data. These equations were shown to be tractable numerically starting with the seminal work \cite{Rattazzi:2008pe} and leading up to \cite{Paulos:2014vya, Simmons-Duffin:2015qma, Behan:2016dtz,  Nakayama:2016jhq, Go:2019lke}, the main tools for the results of this paper. 

The first important step in the numerical conformal bootstrap program \cite{Rattazzi:2008pe} is in focusing on a certain region of the two dimensional space of conformal cross-ratios, now known as the space-like diamond,  in which the conformal blocks ${G_\calO(u,v)}$ turn out to be real and analytic and hence  have Taylor expansions. Equation \eqref{bse} is a vector equation in a real analytic function space. This function space is co-ordinatised as follows \cite{Rattazzi:2008pe} and this is inspired by the knowledge of how \eqref{bse} is solved in the free scalar CFT. There is a  certain point where the convergence of the infinite sums in \eqref{bse} is fastest (for the free scalar CFT): $z = \frac12, \overline{z} = \frac12$, the crossing-symmetric point. The $z, \overline{z}$ co-ordinates for the space of conformal cross-ratios  are related to the original co-ordinates $u,v$ by $u = z \, \overline{z}, ~ v = (1 - z) (1 - \overline{z})$.  Every element of the real analytic function space in which the conformal blocks live, is parametrised by the co-efficients in it's Taylor expansion around this  $z = \frac12, \overline{z} = \frac12$ point, an infinite tuple of real numbers. Thus, the values of a real analytic function and the values of it's various mixed partial derivatives at this special point form the co-ordinates for this function space. Now, one solves equation \eqref{bse} by truncating to  a finite number of co-ordinates, depending on the accuracy needed. The truncation is specified by the maximum number of derivatives retained (both w.r.t $z$ and $\overline{z}$). In obtaining the results of this paper, we have used the semidefinite program solver \cite{Simmons-Duffin:2015qma} via the  \texttt{PyCFTBoot} wrapper \cite{Behan:2016dtz}  to implement the numerical conformal bootstrap. Other resources to implementing the numerical bootstrap can be found in \cite{Paulos:2014vya, Nakayama:2016jhq, Go:2019lke}.

For the last part of  CFT review, we  recall two results, that we will need, and which have already been obtained via the numerical conformal bootstrap.  Both originate from the crossing symmetry constraints on the four point function $\langle \phi \phi \phi \phi \rangle$ of a single scalar operator. 

(i) The first result is what we will refer to as the ``\underline{RRTV bound}'' In \cite{Rattazzi:2008pe}, it was shown that the bootstrap equation requires that the operator in the $\phi \times \phi$ OPE with the lowest scaling dimension is always a scalar, referred to as the lowest scalar, and it's  scaling dimension is bounded from above. This bound is a function of the scaling dimension of $\phi$  ($\Delta_\phi$) and is determined numerically in all dimensions. 

(ii) The second result is what we will refer to as the ``\underline{Rychkov bound}'' \cite{Rychkov:2011et}.  This is a bound on the scalar operator in the $\phi \times \phi$ OPE with next to lowest scaling dimension, if present, which unlike the operator with the lowest scaling dimension need not be a scalar.  The result is that there is an upper bound, determined numerically, for any set of values for $\Delta_\phi$ and for the scaling dimension of the lowest scalar (which is already constrained by the RRTV bound).

\section{The scaling dimension of the lowest scalar}

In this paper, we study a class of CFTs whose low-lying operator spectrum includes four relevant scalars $\sigma$, $\epsilon$, $\sigma'$, $\epsilon'$ in the increasing order of scaling dimensions.   The two dimensional tricritical Ising model \cite{Belavin:1984vu}, \cite{DiFrancesco:1997nk} is one example; there the $\sigma$'s are the $\mathbf{Z}_2$-odd operators while the $\epsilon$'s are the $\mathbf{Z}_2$-even operators. In the $\phi^6$ realisation, the four relevant scalars as $\sigma \leftrightarrow \phi, \epsilon \leftrightarrow \phi^2, \sigma' \leftrightarrow \phi^3, \epsilon' \leftrightarrow \phi^4$. There are  ten OPEs that concern these four fields but we will focus only on two of them viz. the $\sigma \times \sigma$ and  $\epsilon \times \epsilon$ OPEs; since the four point functions whose bootstrap constraints we focus on in this paper, viz. $\langle \sigma \sigma \sigma \sigma \rangle$ and $\langle \epsilon \epsilon \epsilon \epsilon \rangle$ only need these two OPEs for their evaluation. Furthermore, the class of CFT's we study are those with the following particular OPEs:
 \begin{eqnarray} \label{5}
\sigma \times \sigma = \mathbf{1} + \epsilon + \epsilon' + \ldots, \qquad \epsilon \times \epsilon= \mathbf{1} + \epsilon' + \ldots
\end{eqnarray}
That is, the global conformal families of $\epsilon$ and $\epsilon'$ are present in the $\sigma \times \sigma$ OPE and  the global conformal family of $\epsilon'$ is present in the $\epsilon \times \epsilon$ OPE. $\epsilon$ is the scalar operator with the lowest scaling dimension that is present in the $\sigma \times \sigma$ OPE and will be referred to as the `lowest scalar' of the OPE; similarly $\epsilon'$ is the `next to lowest scalar' of the OPE. Also note that $\epsilon'$ is the lowest scalar of the $\epsilon \times \epsilon$ OPE. Thus $\epsilon'$ is the lowest scalar in one OPE and the next to lowest scalar in another.  This is the problem we study here using the numerical conformal bootstrap : CFTs (in all dimensions, 2 to 3 and beyond) with a low lying spectrum and OPE's given by \eqref{5} \cite{Kramers-Wannier} and we will  find that the bootstrap constraints reflect many aspects of tricritical phenomena. 

In this paper, we will work only with unitary CFTs  and use the unitary conformal bootstrap for our analysis. Although it is known that CFTs in fractional dimensions are non-unitary, a fact which was first encountered in \cite{Hogervorst:2014rta} and further expounded in \cite{Hogervorst:2015akt} (see also \cite{Golden:2014oqa}).  As was shown in \cite{Hogervorst:2014rta}, non-unitarity manifests itself in the form of existence of negative norm states at high dimensions.  To quote: ``$\ldots$ a few negative norm states at high dimensions, hidden among lots of positive-norm states of comparable dimensions, probably do not have a strong effect on the low-energy physics. In a recent conformal bootstrap study of the Wilson-Fisher fixed point in fractional dimensions \cite{El-Showk:2013nia},  it was assumed that these theories were unitary, and very reasonable results were obtained $\ldots$'' In the present paper, since we are studying low dimensional operators (the relevant operators), much like \cite{El-Showk:2013nia}, we expect that non-unitarity can be ignored and hence proceed with the unitary bootstrap analysis. 

\subsection{A study in $d = 2$}

We first present an analysis of the crossing symmetry constraints  in two dimensions. We will derive lessons from two dimensions and apply them to other dimensions in the next subsection. Our analysis below gives upper bounds on the scaling dimension of the operator $\epsilon'$ for the class of CFTs under study.

\begin{figure}
  \begin{center} 
  \includegraphics[scale=0.6]{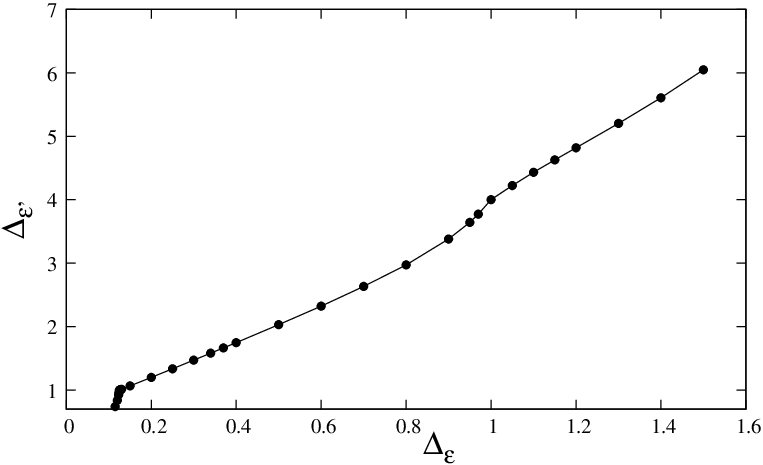}
  \caption{The first upper bound on the scaling dimension of $\epsilon'$: the RRTV bound.
    \label{fig:rrtv} 
  }
  \end{center}
\end{figure}

\underline{The first upper bound}: We obtain an upper bound on the scaling dimension of $\epsilon'$ as follows.  For a given value of $\Delta_\epsilon$, say $s$, we first find the RRTV bound on the lowest scalar in the $\epsilon \times \epsilon$ OPE which is a bound on $\Delta_{\epsilon'}$, $r_1(s)$. We plot the points $(s, r_1(s))$ in a figure where the x-axis is $\Delta_{\epsilon}$ and the y-axis is $\Delta_{\epsilon'}$. We refer to this graph as the RRTV bound on $\Delta_{\epsilon'}$.  For two dimensions it is shown  in figure \ref{fig:rrtv}.

\begin{figure}
\begin{center} 
  \includegraphics[scale=0.6]{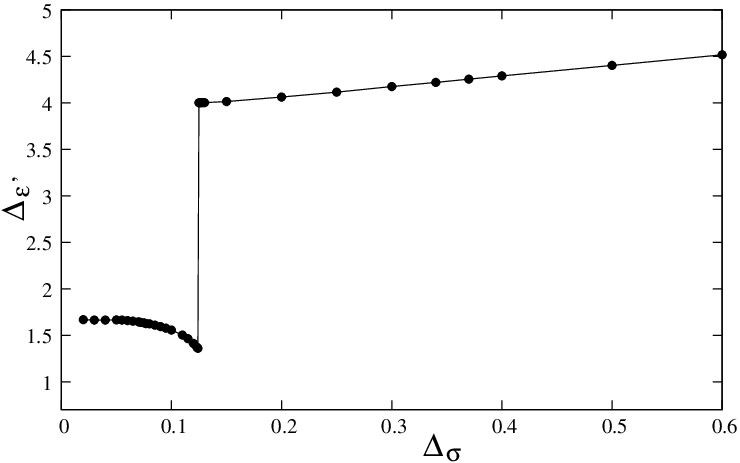}
  \caption{The second upper bound on the scaling dimension of $\epsilon'$: the Rychkov bound.
    \label{fig:rychkov} 
  }
  \end{center}
\end{figure}

\underline{The second upper bound}: We can get another  upper bound on the scaling dimension of $\epsilon'$ in the following way.  For a given value of $\Delta_\sigma$, say $r$, we first find the RRTV bound on the lowest scalar in the $\sigma \times \sigma$ OPE, which is a bound on $\Delta_\epsilon$,  $r_2(r)$.  Then for the pair $(\Delta_\sigma = r, \Delta_\epsilon = r_2(r))$ we find the Rychkov bound on the next to lowest scalar in the $\sigma \times \sigma$ OPE which is a bound on $\Delta_{\epsilon'}$,  $r_3(r)$. We plot the points $(r, r_3(r))$ in a figure where the x-axis is $\Delta_{\sigma}$ and the y-axis is $\Delta_{\epsilon'}$.  We  refer to this graph as the Rychkov bound on $\Delta_{\epsilon'}$. For two dimensions it is shown in figure \ref{fig:rychkov}.

\begin{figure}
  \begin{center} 
  \includegraphics[scale=0.6]{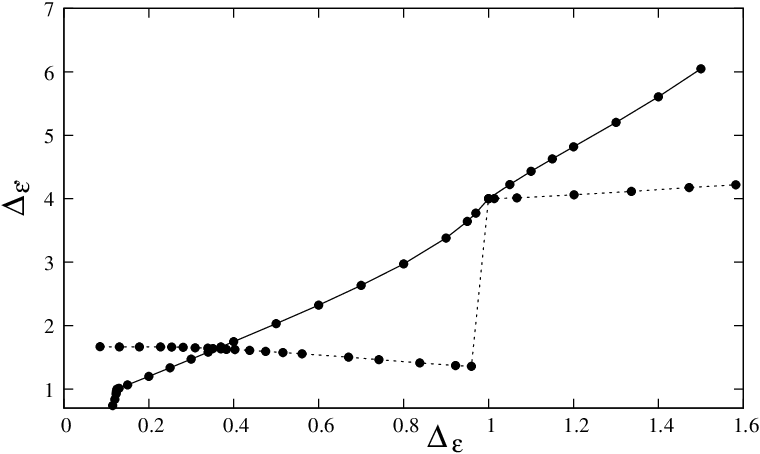}
  \caption{The RRTV (solid) and the Rychkov (dashed) bounds in two dimensions. The two bounds cross at $r_2(0.075 \pm 0.001$).
    \label{fig:one} 
  }
  \end{center}
\end{figure}

Now we will compare the RRTV and Rychkov bounds on $\Delta_{\epsilon'}$. First we note that the RRTV bound in figure \ref{fig:rrtv} is a $\Delta_{\epsilon}-\Delta_{\epsilon'}$ plot while the Rychkov bound in figure \ref{fig:rychkov} is a $\Delta_{\sigma}-\Delta_{\epsilon'}$ plot. We can convert the latter also into a $\Delta_{\epsilon}-\Delta_{\epsilon'}$ plot as follows:  we plot the points $(r_2(r), r_3(r))$ instead of $(r, r_3(r))$.  We can now put the RRTV and Rychkov bounds in the same $\Delta_{\epsilon}-\Delta_{\epsilon'}$ plot by plotting the points $(s, r_1(s))$ and $(r_2(r), r_3(r))$ respectively. This is shown for two dimensions in figure \ref{fig:one}.

From figure \ref{fig:one} we observe that for smaller values of  $\Delta_\sigma$ ($= r$)  (which corresponds to smaller values of $r_2(r)$, because $r_2(r)$ is a monotonically increasing function), we find that the Rychkov bound is bigger than the RRTV bound.  For subsequent larger values of $\Delta_\sigma$, the graph for the Rychkov bound  crosses the graph for the RRTV bound so that to the left of the crossing point the Rychkov bound is larger than the RRTV bound and to the right of the crossing point the RRTV bound is larger.  There is a specific value for $\Delta_\sigma$, say $\Delta_\sigma^{\text{cross}}$, for which this crossing happens; to be precise  $r_2(\Delta_\sigma^{\text{cross}})$ is the value on the x-axis for the crossing point.  In two dimensions, the graphs for the RRTV and Rychkov bounds are plotted in figure \ref{fig:one} and we find $\Delta_{\sigma}^{\text{cross}} = 0.075 \pm 0.001$. This compares well with  the known value for $\Delta_\sigma$ in the two dimensional tricritical Ising CFT (which is  $0.075$) within error bars.  Hence, we conclude that the value of $\Delta_{\sigma}^{\text{cross}}$, determined by conformal bootstrap constraints as described above, is the value of $\Delta_\sigma$ in the CFT.

A more succinct description of the analysis of this subsection can be made as follows \cite{thanxtoreferee}.  $\Delta_{\epsilon'}$ can be maximised keeping $\Delta_\epsilon$ fixed from the $\langle\epsilon \epsilon \epsilon \epsilon\rangle$ four point function. Alternatively $\Delta_{\epsilon'}$ can be maximised in the $\langle \sigma \sigma \sigma \sigma \rangle$ four point function keeping also $\Delta_\epsilon$ to the maximal allowed value. Then the observation is that the tricritical Ising CFT in two dimensions satisfies these two maximisation conditions simultaneously. 

\underline{Some details pertaining to the numerical study}:  We have used the program \texttt{PyCFTBoot} \cite{Behan:2016dtz} which is a program written in Python and which is based on the \texttt{SDPB} solver \cite{Simmons-Duffin:2015qma}; these resources give us access to the semidefinite programming methods for the bootstrap pioneered in \cite{Poland:2011ey, Kos:2013tga, Kos:2014bka}. In all the computations  that we are reporting in this paper and the plots we have displayed, we have used the following user-defined input parameters in \texttt{PyCFTBoot}: $k_{max} = 30, l_{max} = 30, m_{max} = 7, n_{max} = 10.$ This sets up a table associated with conformal blocks (\texttt{ConformalBlockTable} in \texttt{PyCFTBoot}) with $30$ poles, $30$ spins and a $7 \times 10$ triangle of derivatives (which amounts to working with $m_{max} + 2\, n_{max} = 27$ derivatives of the conformal blocks).  We did even compute with up to $40$ poles, $40$ spins and a $8 \times 12$ triangle of derivatives (32 derivatives). This increases the computing time substantially without adding any substantial improvement in the details beyond what we are reporting here. In the \texttt{SDPB} command used to bisect over gaps in a scalar operator \texttt{sdp.bisect(lower, upper, tol, 0)} we have changed from the  default tolerance value of $\texttt{tol}=0.01$ to a tolerance value of $\texttt{tol}=0.00001$. This was essential to improve the precision of the Rychkov bound.

\subsection{ $2 < d < 3$ and beyond}

\begin{figure}
     \centering
     \begin{subfigure}[b]{0.23\textwidth}
         \centering
         \includegraphics[width=\textwidth]{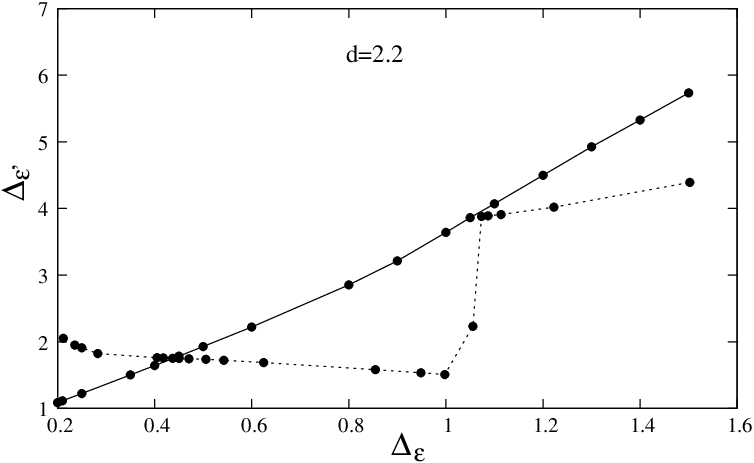}
         \caption{}
         \label{fig:443}
     \end{subfigure}
     \begin{subfigure}[b]{0.23\textwidth}
         \centering
         \includegraphics[width=\textwidth]{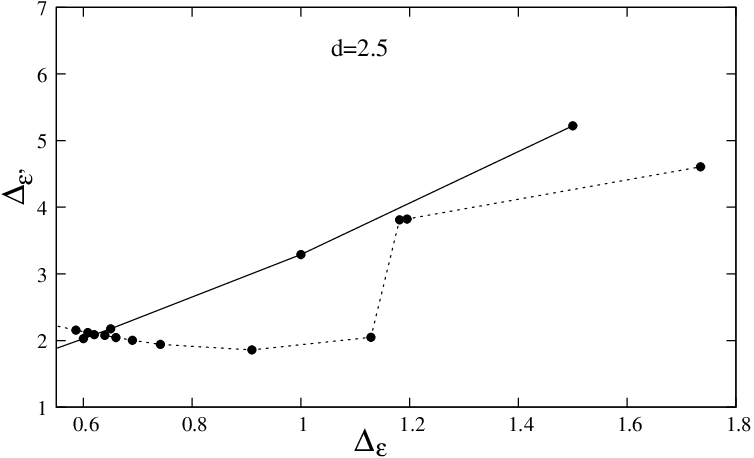}
         \caption{}
         \label{fig:444}
     \end{subfigure}
     \hfill
     \begin{subfigure}[b]{0.23\textwidth}
         \centering
         \includegraphics[width=\textwidth]{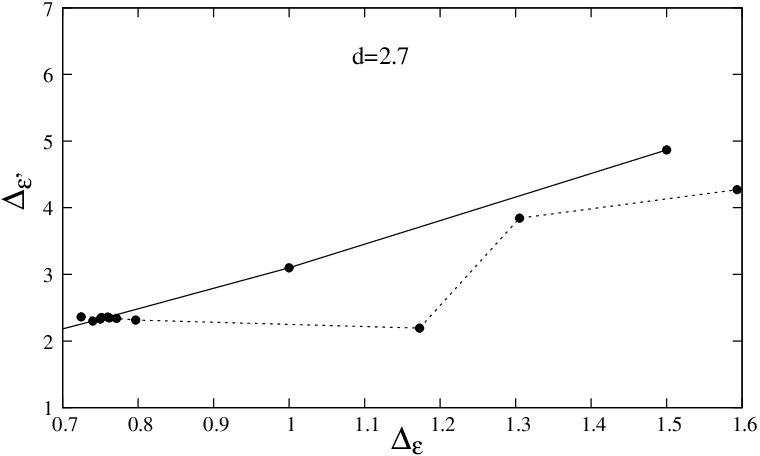}
         \caption{}
         \label{fig:449}
     \end{subfigure}
     \hfill
     \begin{subfigure}[b]{0.23\textwidth}
         \centering
         \includegraphics[width=\textwidth]{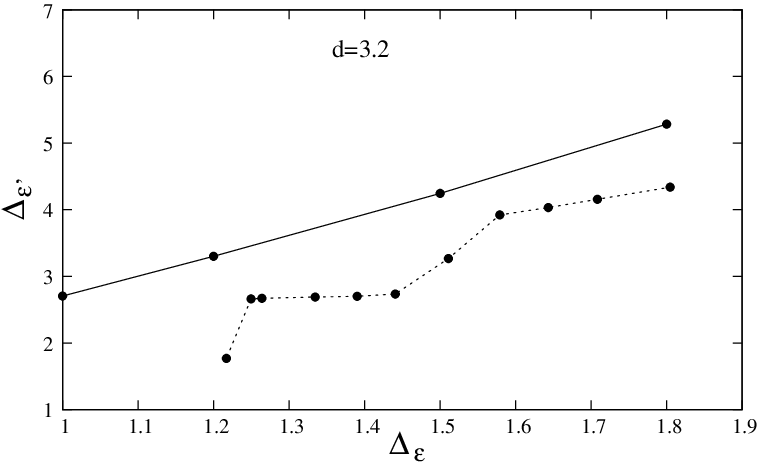}
         \caption{}
         \label{fig:4410}
     \end{subfigure}
     \hfill
        \caption{The RRTV (solid) and the Rychkov (dashed) bounds in various non-integer dimensions: $(a)\, d=2.2$, $(b)\, d=2.5$, $(c)\, d=2.7$, $(d)\, d=3.2$. The two bounds cross at  $(a)\,r_2(0.136)$,  $(b)\,r_2(0.259)$,  $(c)\,r_2(0.352)$ and for $(d)$ there is no crossing. }
        \label{nid}
\end{figure}

\begin{figure}[ht]
  \begin{center} 
  \includegraphics[scale=0.5]{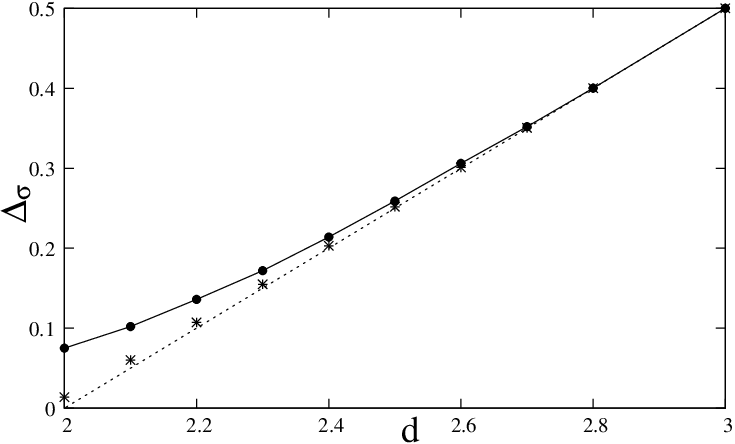}
  \caption{The bold line shows the CFT computation for $\Delta_\sigma$, i.e. $\Delta_\sigma^{\text{cross}}$ and the dashed line shows the unitarity bound for scalar operators, for $2 \leq d \leq 3$.  As the spatial dimension increases $\Delta_\sigma^{\text{cross}}$ becomes very close to the unitarity bound. Points represented by $\ast$ and which are not joined into a line represent the $\epsilon$-expansion computation for $\Delta_\sigma$.
   \label{fig:smn} 
  }
  \end{center}
\end{figure}

The equations that encode the bootstrap constraints and in fact the whole formalism is analytic in the number of dimensions $d$  and thus provides a remarkable way to study CFTs in non-integer dimensions.  The Ising model CFT  was studied in all dimensions $2 \leq d \leq 4$ in \cite{El-Showk:2013nia}. Here, we study our class of tricritical Ising CFTs in a generic non-integer dimension $d$. We numerically determine the RRTV and Rychkov bounds and look for the point, if any, where the bounds coincide and from that obtain $\Delta_\sigma^{\text{cross}}$ in the non-integer dimension $d$. The results of this investigation are presented in three ways. First, in the figure  \ref{nid}, for four different representative non-integer dimensions $d = 2.2, 2.5, 2.7, 3.2$, we plot the RRTV and Rychkov bounds and the point of intersection which gives the $\Delta_\sigma^{\text{cross}}$.  Second, in the table \ref{tab:a}, in the first two columns, we give the $\Delta_\sigma^{\text{cross}}$ for  more values of $d$ in the range $2 < d < 3$ where there is a crossing or  intersection of the two bounds. Third, in the plot of figure \ref{fig:smn}, we show a comparison between the CFT computation $\Delta_\sigma^{\text{cross}}$ with the unitarity bound for scalar operators in the dimension range $2 < d < 3$.

First we will discuss the range of dimensions $2 < d < 3$ where there is a crossing of bounds. We find that $\Delta_\sigma^{\text{cross}}$ increases as $d$ increases from $2$ to $3$ (see table \ref{tab:a} as well as figure \ref{fig:smn}). Note that $\Delta_\sigma^{\text{cross}}$ is bigger than the unitarity bound, $\frac{d-2}{2}$,  for dimensions between $2$ and $3$ and this difference is maximum for $d = 2$ and decreases with increasing $d$. We also  find that  $\Delta_\sigma^{\text{cross}}$ approaches $\frac12$ as $d$ approaches $3$. This matches with the known value for $\Delta_\sigma$ in $d=3$: the upper critical dimension for the tricritical Ising model  is $3$ wherein the scaling dimension of $\sigma$ is it's classical value of $\frac12$. The numerical computation  becomes harder as $d \rightarrow 3$ because $\Delta_\sigma^{\text{cross}}$ seems to be very close to the unitarity bound. 

Our numerical studies for $3  <  d <  4$ show that the two bounds do not cross. See the fourth plot  in figure \ref{nid}. This observation is consistent with known facts. The tricritical Ising exponents for $3  <  d <  4$ should be the same as for three dimensions which is a violation of the unitarity bound (for $\Delta_\sigma$) and hence one is in the realm of non-unitary CFTs. But our analysis, following\, \cite{Rattazzi:2008pe, Rychkov:2011et} is an analysis of constraints on unitary CFTs.

Our surmise that $\Delta_\sigma^{\text{cross}}$ gives the value for $\Delta_\sigma$ in the CFT thus has passed non-trivial tests by reproducing the known values in $d = 2$, $d = 3$ and in $3 < d < 4$ and hence the computations for $2 < d < 3$ constitute a prediction coming from conformal bootstrap analysis. 

\begin{table} 
\caption{$\Delta_\sigma$ in dimensions $2$ to $3$ from CFT analysis and from $\epsilon$-expansion.}
\label{tab:a}
	\centering
		\begin{tabular}{|l|l|l|}  
\hline
		Dimension (d) &  $\Delta_\sigma^{\text{cross}}$ from CFT  & $\Delta_\sigma$ from $\epsilon$-expansion\\
\hline
$3.00$ & $-$ & $0.50000$  \\		\hline
$2.90$ & $-$ & $0.45002$\\	\hline
$2.80$ & $0.4002(1)$ & $0.40014$\\	\hline
$2.70$  & $0.352(1)$ & $0.35043$\\	\hline
$2.60$ & $0.306(1)$ & $0.30098$\\	\hline
$2.50$ & $0.259(1)$ & $0.25184$\\	\hline
$2.40$ & $0.214(1)$ & $0.20311$\\	\hline
$2.30$ & $0.172(1)$ & $0.15486$\\	\hline
$2.20$ & $0.136(1)$ & $0.10716$\\	\hline
$2.10$ & $0.102(1)$ & $0.06010$\\	\hline
$2.00$ & $0.075(1)$ & $0.01374$\\	\hline
		\end{tabular}
\end{table}

\subsection{Comparison with $\epsilon$-expansion}

Before the advent of bootstrap methods, $\epsilon$-expansion techniques were used to study CFTs in non-integer dimensions. The CFTs associated with the critical Ising model have been studied extensively in this way, starting with the seminal works of Wilson and Fisher \cite{Wilson:1971dc}, \cite{Wilson:1972cf}.  For the Ising model ($\phi^4$ theory), $\epsilon$-expansion results up to a very high order ($\epsilon^5$) are available \cite{LeGuillou:1987ph}, \cite{Guida:1998bx}, \cite{Kleinert:1991rg} (see also \cite{ZinnJustin:2002ru}, \cite{Kleinert:2001ax}). But the $\epsilon$-expansion results for tricritical CFTs ($\phi^6$ theory) are not available to that high an order; computations up to $\epsilon^3$-order were done in\,\cite{Lewis:1978zz, Hager:2002uq}. From \cite{equationinHager}, we obtain the $\epsilon$-expansion up to order three for the critical exponent $\eta$ and further using the relation between the critical exponent $\eta$ and the scaling dimension of the lowest scalar viz. 
\begin{equation}
\eta = 2 - d + 2\, \Delta_{\sigma}
\end{equation}
we have the following $\epsilon$-expansion result ($\epsilon = 3 - d$):  
\begin{equation} \label{sigmaepsilon}
\Delta_{\sigma} = \frac{1}{2}-\frac{\epsilon}{2}  + \frac{\epsilon^2}{1\,000} + \frac{10125\, \pi^2 + 91160}{15\,000\,000} \epsilon^3 + \ldots
\end{equation}

We now test our hypothesis by comparing the values of $\Delta_{\sigma}^{\text{cross}}$ obtained above with the best known values from $\epsilon$-expansion (equation \eqref{sigmaepsilon}). The comparison between the CFT results and $\epsilon$-expansion results is shown in table \ref{tab:a}.  It is also shown in figure \ref{fig:smn} where the CFT values are joined into a line and the $\epsilon$-expansion values are points $\ast$ not joined into a line. The results coincide in dimensions close to three, but  deviate as one approaches two dimensions.  This is perhaps because the study of the tricritical Ising point using $\epsilon$-expansion is known to give poor estimates in two dimensions. One expects that modern approaches to $\epsilon$-expansion computations that incorporate conformal symmetry such as the Rychkov-Tan\,\cite{Rychkov:2015naa} method and the Polyakov-Mellin bootstrap\,\cite{Gopakumar:2016wkt} would give better results \cite{wp1}. 

\section{Co-existence of multiple universality classes and Multicriticality}

\underline{Two Plateaus} : 
\newline In the previous section, we started our discussion with a bootstrap analysis in two dimensions. Together with exact results available in two dimensions, we gathered some lessons from the two dimensional analysis and then applied them to dimensions other than two. Here in this section, we will now again focus on another observation about two dimensions. In the plot of the two dimensional Rychkov bound (dashed line of figure \ref{fig:one}), we see that there are two plateaus.  One is the plateau that starts around $\Delta_{\epsilon}=1, \Delta_{\epsilon'}=4$.  This one has been well studied in\,\cite{Rychkov:2011et}; it was argued that within this $\Delta_\epsilon-\Delta_{\epsilon'}$ space, as one approaches the Ising CFT, the operator  $\epsilon'$ becomes irrelevant (in fact, it was used as one of the criteria to partially isolate the Ising CFT). Hence this plateau can be associated with the Ising universality class.  

There is a second plateau in figure \ref{fig:one}:  through out this plateau $\Delta_{\epsilon} < 1$  and $\Delta_{\epsilon'} < 2$.  If a certain CFT is such that $\epsilon'$ can not be an irrelevant operator in it (such as the two dimensional tricritical Ising CFT) then that CFT has to exist in the region ($\Delta_{\epsilon} < 1$) of this second plateau.   We can thus associate this second plateau with the tricritical Ising universality class. 

The two plateaus in the plot of the Rychkov bound indicate that the conformal bootstrap constraints allow for the (possible) existence of two different universality classes: one where $\epsilon'$ is relevant and another where $\epsilon'$ is irrelevant or in other words the Ising and the tricritical Ising universality  classes. This result from CFT analysis is consistent with our understanding  from studies of $\phi^6$ field theory\,\cite{lawrie}: the $\epsilon'$ operator is the $\phi^4$ operator and it is known that the $\phi^4$ operator determines whether the flow is towards the tricritical point or the Ising point. 

\begin{figure}
\centering
\includegraphics[scale =0.5]{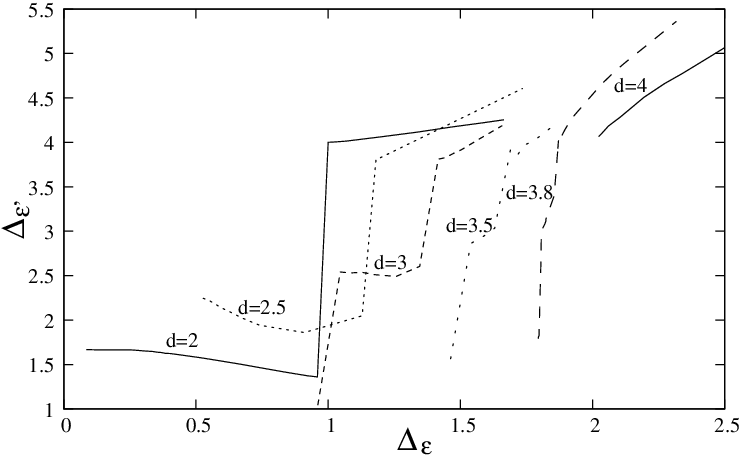}
\caption{The plots of the Rychkov bound in various non-integer dimensions, consist of two plateaus. The higher plateau is associated with the Ising universality class while the lower one with the tricritical Ising universality class. The lower plateau is smaller for higher dimensions and vanishes at four dimensions.  }
\label{fig3}
\end{figure}

Thus motivated by the above observations on plateaus in two dimensions we plot the Rychkov bound in various non-integer dimensions in figure \ref{fig3} and find that the two plateau structure exists there too.  This is the conformal bootstrap signature for the existence of both the Ising and tricritical Ising universality classes in non-integer dimensions as well.  On closer examination, we find that the width of the lower plateau decreases with increasing dimension and vanishes at $d=4$.

\begin{figure}
\begin{tabular}{cc}
\includegraphics[width=4 cm]{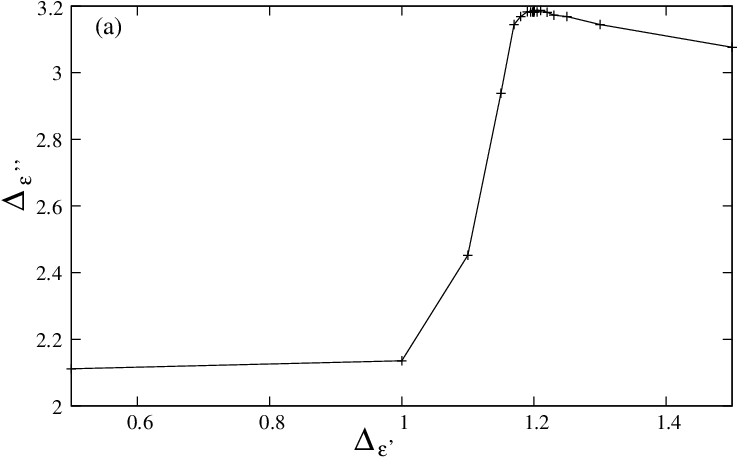}
& \includegraphics[width=4 cm]{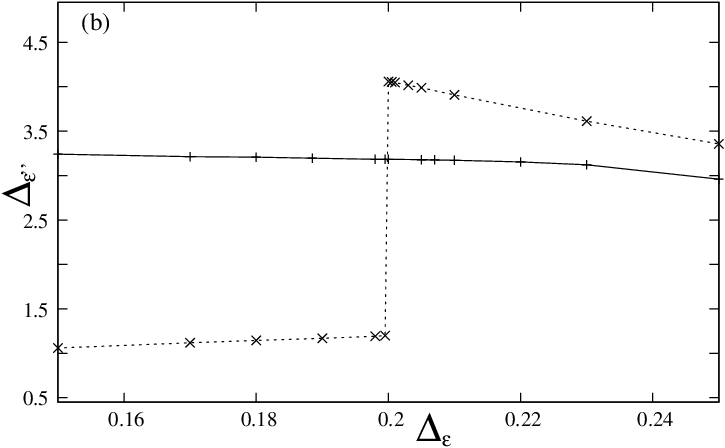}\\
\end{tabular}
\caption{a) Upper bound on $\Delta_{\epsilon''}$ from the bootstrap constraints associated to the $\langle \sigma \sigma \sigma \sigma \rangle$ correlator  in two dimensions for $\Delta_\sigma = 0. 075,  \Delta_\epsilon = 0.2$; b)  Upper bound on $\Delta_{\epsilon''}$ from the bootstrap constraints associated to the $\langle \sigma \sigma \sigma \sigma \rangle$ correlator (solid line) and the $\langle \epsilon \epsilon \epsilon \epsilon \rangle$ correlator (dashed line) in two dimensions for $\Delta_\sigma = 0. 075,  \Delta_\epsilon = 0.2$. 
\label{fig:sm}
}
\end{figure}

\underline{The $\epsilon''$ Operator} :
\newline  So far, we have been working with the class of CFTs consisting of $\sigma, \epsilon, \epsilon'$ operators in the lowest part of the spectrum with OPEs \eqref{5}. Here, we will introduce into our considerations the operator above these, which is denoted by $\epsilon''$ and the OPEs are now 
 \begin{eqnarray} \label{8}
\sigma \times \sigma = \mathbf{1} + \epsilon + \epsilon' + \epsilon'' +  \ldots, \qquad \epsilon \times \epsilon= \mathbf{1} + \epsilon' + \ldots.
\end{eqnarray}
The two dimensional tricritical Ising CFT is an example of this class of CFTs (from which the notation is derived) where $\epsilon''$ is the primary operator with the largest scalaing dimension. In the $\phi^6$ realization, $\epsilon'' \leftrightarrow \phi^6$. 

We will now analyse the bootstrap constraints for this class of CFTs.  In two dimensions,  when we fix  $\Delta_\sigma$ and $\Delta_\epsilon$ to their exact values of $0.075$ and $0.2$ respectively, the bootstrap equations from the $\langle \sigma \sigma \sigma \sigma \rangle$ correlator give an upper bound on $\Delta_{\epsilon''}$  for a given value of $\Delta_{\epsilon'}.$  This bound on $\Delta_{\epsilon''}$ as a function of $\Delta_{\epsilon'}$ is plotted in figure \ref{fig:sm}(a) and one observes that the bound on $\Delta_{\epsilon''}$   shows a discontinuous jump to $ \approx 3.1$ at $\Delta_{\epsilon'}=1.2$. Noting that the minimal model values for $\epsilon''$ and $\epsilon'$ are $3.0$ and $1.2$ respectively, this discontinuous jump occurs close to the exact values.  

In another computation, we fix $\Delta_\sigma$ and $\Delta_{\epsilon'}$ to their minimal model values.  Constraints from the $\langle \sigma \sigma \sigma \sigma \rangle$ and the $\langle \epsilon \epsilon \epsilon \epsilon \rangle$ correlators give upper bounds on $\Delta_{\epsilon''}$ as a function of $\Delta_{\epsilon}$ which are plotted in figure \ref{fig:sm}(b) as the solid and dotted lines respectively. When one traces the lower of the two upper bounds, one finds a jump at $\Delta_\epsilon = 0.2$ where $\Delta_{\epsilon''}$ is $\approx 3.1$. Again these are very close to the exact minimal model values.  

Thus we show that in two dimensions, the tricritical Ising CFT seems to be characterised by the property that $\epsilon''$ goes to being irrelevant as a function of $\Delta_{\epsilon'}$, while keeping both $\Delta_{\sigma}$ and $\Delta_{\epsilon}$ fixed. Much like the Ising CFT \cite{Rychkov:2011et} which was characterised by the property that $\epsilon'$ goes from being relevant to irrelevant, as a function of $\Delta_{\epsilon}$, while keeping $\Delta_{\sigma}$ fixed. Hence we conclude that the tricritical Ising CFT could also be isolated by using a bootstrap analysis much like the Ising CFT.  The analysis so far, involving the $\epsilon''$ operator, has been in two dimensions only where exact results are available. Isolating the tricritical Ising CFT completely in dimensions beyond two, using considerations from the $\epsilon''$ operator, would constitute progress.

\underline{Multicriticality} :
\newline In this section, so far we have (mostly) studied the bootstrap constraints from only one four point function viz. the pure $\langle \sigma \sigma \sigma \sigma \rangle$ correlator and this involves only the subset of the CFT data that occur in the $\sigma \times \sigma$ OPE.   We will now show that these constraints could encode even more information, even of higher order critical points. First, to reiterate, we have seen how  the Ising CFT is characterised by the property that $\epsilon'$ goes from being relevant to irrelevant, as a function of $\Delta_{\epsilon}$, while keeping $\Delta_{\sigma}$ fixed . We have also shown that the tricritical Ising CFT is characterised by the property that $\epsilon''$ goes to being irrelevant as a function of $\Delta_{\epsilon'}$, keeping both $\Delta_{\sigma}$ and $\Delta_{\epsilon}$ fixed.  Going on, one introduces the next operator viz. $\epsilon'''$ and studies the class of CFTs with the OPE $\sigma \times \sigma = \mathbf{1} + \epsilon + \epsilon' + \epsilon'' +  \epsilon''' +  \ldots$  The next higher order critical point would be where the $\Delta_{\epsilon'''}$  would go from relevant to irrelevant as a function of $\Delta_{\epsilon''}$, keeping all of $\Delta_{\sigma}, \Delta_{\epsilon}$ and $\Delta_{\epsilon'}$ fixed. We hope to report investigations along these lines in a future work \cite{wp1}. Other studies of critical and multicritical models using CFT and other methods include \cite{Codello:2017qek, Codello:2017hhh, Codello:2017epp}.

\section{Conclusion}

In conclusion, we have shown that the tricritical point, which is known to be unstable because the system can easily crossover to the ordinary critical point, can be studied using conformal bootstrap techniques. Such non-perturbative CFT methods are even more significant for the tricritical point, as the success with $\epsilon$-expansion has been limited.  Using the bootstrap constraints coming from only two pure correlators on the CFT data contained in two OPEs, we have seen many signatures of tricritical physics and also obtained the precise value of one critical exponent.  To get  precise values for the other exponents one will have to consider crossing symmetry constraints coming from all correlators, pure and mixed,  which would involve CFT data in other OPEs  \cite{wp1}.


\begin{thebibliography}{99}
\bibitem{equationinHager} This is essentially equation (23) of reference \cite{Hager:2002uq} which gives the result for $O(n)$ models and for the tricritical Ising case that we are interested in, we need to set $n=1$ in that equation.
\end{thebibliography}
\end{document}